\documentclass{aa}  
\usepackage{graphicx}
\usepackage{txfonts}
\usepackage{natbib}
\bibpunct{(}{)}{;}{a}{}{,}
\def\ltapprox{\raise 2pt \hbox {$<$} \kern-1.1em \lower 5pt \hbox {$\approx$}}
\def\ltsim{\raise 2pt \hbox {$<$} \kern-1.1em \lower 4pt \hbox {$\sim$}}
\def\gtsim{\raise 2pt \hbox {$>$} \kern-1.1em \lower 4pt \hbox {$\sim$}}
\begin{document}
   \title{Mass distribution in the most X-ray-luminous galaxy cluster \\
RX J1347.5$-$1145 studied with \textit{XMM--Newton}}

   \subtitle{}

   \author{M. Gitti
          \inst{1}
          \and
          R. Piffaretti\inst{2}
          \and
          S. Schindler\inst{3}\fnmsep
          }

   \offprints{M. Gitti}

   \institute{INAF - Osservatorio Astronomico di Bologna,
              via Ranzani 1, I-40127 Bologna, Italy\\
              \email{myriam.gitti@oabo.inaf.it}
         \and 
             SISSA/ISAS, via Beirut 4, I-34014 Trieste, Italy
         \and 
             Institut f\"ur Astro- und Teilchen Physik,
              Leopold-Franzens Universit\"at Innsbruck, 
              Technikerstra\ss e 25, A-6020 Innsbruck, Austria    
             }

\authorrunning{Gitti et al.}
\titlerunning{Mass distribution in RX J1347.5$-$1145 studied with \textit{XMM--Newton}}

   \date{Received ; accepted }


\abstract
{We report on the analysis of \textit{XMM-Newton} observations of RX
  J1347.5$-$1145 (z=0.451), the most X-ray-luminous galaxy cluster.}
{We present a detailed total and gas mass determination up to large
  distances ($\sim$ 1.7 Mpc), study the scaling properties of the
  cluster, and explore the role of AGN heating in the cluster cool core.
}
{By means of spatially resolved spectroscopy we derive density,
  temperature, entropy, and cooling time profiles of the intra-cluster
  medium. We compute the total mass profile of the cluster in the
  assumption of hydrostatic equilibrium.  
}
{If the disturbed south-east region of the cluster is excluded from
  the analysis, our results on shape, normalization, scaling
  properties of density, temperature, entropy, and cooling time
  profiles are fully consistent with those of relaxed, cool core
  clusters. We compare our total and gas mass estimates
  with previous X-ray, lensing, dynamical, and SZ studies. We find
  good agreement with other X-ray results, dynamical mass
  measurements, weak lensing masses and SZ results.  We confirm a
  discrepancy of a factor $\sim$2 between strong lensing and X-ray
  mass determinations and find a gross mismatch between our total mass
  estimate and the mass reconstructed through the combination of both
  strong and weak lensing. We explore the effervescent heating
  scenario in the core of RX J1347.5$-$1145 and find support to the
  picture that AGN outflows and heat conduction are able to quenching
  radiative cooling.}
   {}

   \keywords{  Galaxies: clusters: individual: RX J1347.5$-$1145 -- 
               X-rays: galaxies: clusters   --
               intergalactic medium --
               cooling flows --
               dark matter --
               Cosmology: observations }

   \maketitle
%


\section{Introduction}

X-ray observations of the diffuse Intra-Cluster Medium (ICM) in
clusters of galaxies are a particularly rich source of information for
understanding the formation of large scale structure and the physics
of clusters. As they are the last manifestation of hierarchical
clustering, whose history depends strongly on cosmology, galaxy
clusters are key objects for cosmological studies \citep[see][for a
review]{voit05}.  Since the evolution of the ICM is mainly driven by
the gravity of the underlying dark matter halo, clusters are expected
to show similar properties when rescaled with respect to their total
mass and formation epoch.  However, deviations from self-similarity
are expected under the effect of more complex physical processes,
beyond gravitational dynamics only, which affect the thermodynamical
properties of the diffuse ICM \citep[e.g.][and references
therein]{evrardhenry91,bryanandnorman,borgani02}.  It is therefore
essential to investigate whether galaxy clusters obey the expected
scaling relations, which are the foundation to use these virialized
objects as cosmological probes.  The first important step in this
context is to find a proxy for an accurate determination of the
cluster mass.

The galaxy cluster RX J1347.5$-$1145 (z=0.451) is an exceptional
object in many aspects.  It is the most X-ray-luminous cluster known
to date ($L_X = 6 \times 10^{45}$ erg s$^{-1}$ in the [2-10] keV
energy range) with a very peaked surface brightness profile and hosts
a strong cooling flow in its center with nominal mass accretion rate
of $\sim 1900~{\rm M}_{\odot} {\rm yr}^{-1}$ \citep{gitti04}.  The
cluster is dominated by two cD galaxies which are separated by about
$\sim 18''$ along the east-west direction, the X-ray emission being
centered on the western one.  Although this is unusual for strong
cooling flow clusters, the optical spectrum of the western Brightest
Cluster Galaxy (BCG) indicates that it hosts an active galactic
nucleus (AGN), with typical emission lines of giant ellipticals at the
center of cooling flow clusters \citep{cohen2002}.  More striking is a
recent discovery made with \textit{Chandra} \citep{allen02b} and
\textit{XMM-Newton} \citep{gitti04} of a region with hot, bright X-ray
emission located at $\sim 20$ arcsec from the central emission peak in
south-east direction.  Submillimeter observations also detected a very
deep SZ decrement in the south-east region of the cluster
\citep{komatzu99,etienne2001}.  These results were interpreted as
indications of a subcluster merger in an otherwise relaxed, massive
cool core cluster, pointing to a complex dynamical evolution of the
system.  Furthermore, RX J1347.5$-$1145 is a powerful gravitational
lens and mass reconstructions based on weak and strong lensing
analyses have been performed
\citep{schindler1995,fischer97,sahu1998,bradac05}.

With a detailed study of the properties of the ICM in this cluster it
is thus possible to address many key issues on both dynamical and
non-gravitational processes in galaxy clusters.  A great advantage of
observing RX J1347.5$-$1145 with \textit{XMM-Newton} is that important
quantities derived for the undisturbed cluster (i.e., with the
south-east quadrant excluded) such as the azimuthally averaged ICM
density and temperature profiles can be computed up to a large
distance from the center ($\sim$ 1730 kpc).  The measurement of
cluster temperature gradients at large distances is also crucial for
determining the total gravitational masses and in turn the gas mass
fraction of clusters.  A precise determination of the total mass at
large radii allows an estimate of the virial radius of the object
without much extrapolation of the universal NFW dark matter profile
\citep{nfw96}.  The virial radius can then be used to study the
scalings of the temperature and entropy profiles and a fair comparison
between predictions of numerical simulations and observations can be
performed.  Currently, the two most promising techniques for obtaining
accurate determinations of cluster masses are X-ray observations, by
deprojection of X-ray surface brightness combined with spectroscopic
determination of the cluster temperature, and gravitational lensing,
through either strong lensing features or statistical distortions of
background objects (weak lensing).  The mass estimates inferred with
these two methods can be quite inconsistent, particularly in the case
of strong lensing \citep[e.g.][and references therein]{wu98}. In
contrast to the X-ray technique, the gravitational lensing method is
essentially free of assumptions on the nature and the dynamical state
of the gravitating material.  In particular, the X-ray method can be
affected strongly during mergers \citep{schindler1996} and in the
inner cluster region where a strong interaction between the central
AGN and the ICM is present \citep[e.g.,][]{birzan}, as in these cases
deviations from the assumptions of hydrostatic equilibrium and
spherical symmetry are expected.  Since both the total mass profile
derived from X-rays and the total mass distribution derived from
gravitational lensing are available for RX J1347.5$-$1145, a
comparison between them is possible thus providing important insights
on this issue.  Furthermore, the presence of gas with short cooling
time in the cluster core offers the opportunity to explore gas heating
processes such as AGN heating, which have become increasingly popular
since the failure of standard cooling flows models.

In this paper, by starting from the results of morphological
(Sect. \ref{morphology.sec}) and spectral (Sect. \ref{spectral.sec})
analyses of \textit{XMM--Newton} observations of RX J1347.5$-$1145
(Sect. \ref{data.sec}), we present a detailed study of the cluster
mass distribution (Sect. \ref{mass.sec}), and discuss its comparison
with the mass profile derived from previous studies
(Sect. \ref{comp.sec}).  We also study the scaling properties of the
cluster (Sect. \ref{radial.sec} and \ref{mass.sec}) and explore the
role of AGN heating in the cluster cool core in the context of the
effervescent heating scenario (Sect. \ref{coolcore.sec}).  RX
J1347.5$-$1145 (hereafter RX J1347) is at a redshift of 0.451.  With a
Hubble constant of $H_0 = 70 \mbox{ km s}^{-1} \mbox{ Mpc}^{-1}$, and
$\Omega_M = 1-\Omega_{\Lambda} = 0.3$, the luminosity distance is 2506
Mpc and the angular scale is 5.77 kpc per arcsec.


\section{Observation and data preparation}
\label{data.sec}

RX J1347 was observed by \textit{XMM--Newton} in July 2002 during rev.
484 with the MOS and pn detectors in Full Frame Mode with THIN filter,
for an exposure time of 37.8 ks for MOS and 33.2 ks for pn.  We use
the SASv6.0.0 processing tasks \textit{emchain} and \textit{epchain}
to generate calibrated event files from raw data.  Throughout this
analysis single pixel events for the pn data (PATTERN 0) are selected,
while for the MOS data sets the PATTERNs 0-12 are used.  The removal
of bright pixels and hot columns is done in a conservative way
applying the expression (FLAG==0).  To reject the soft proton flares
we accumulate the light curve in the [10-12] keV band for MOS and
[12-14] keV band for pn, where the emission is dominated by the
particle-induced background, and exclude all the intervals of exposure
time having a count rate higher than a certain threshold value (the
chosen threshold values are 0.15 cps for MOS and 0.22 cps for pn). The
remaining exposure times after cleaning are 32.2 ks for MOS1, 32.5 ks
for MOS2 and 27.9 ks for pn.  Starting from the output of the SAS
detection source task, we make a visual selection on a wide energy
band MOS \& pn image of point sources in the field of view (hereafter
FOV). Events from these regions are excluded directly from each event
list.

The background estimates are obtained using a blank-sky observation
consisting of several high-latitude pointings with sources removed
\citep{lumb2002}.  The blank-sky background events are selected using
the same selection criteria (such as PATTERN, FLAG, etc.), intensity
filter (for flare rejection) and point source removal used for the
observation events; this yields final exposure times for the blank
fields of 365 ks for MOS1, 350 ks for MOS2 and 294 ks for pn.  Since
the cosmic ray induced background might slightly change with time, we
compute the ratio of the total count rates in the high energy band
([10-12] keV for MOS and [12-14] keV for pn). The obtained
normalization factors (0.992, 1.059, 1.273 for MOS1, MOS2 and pn,
respectively) are then used to renormalize the blank field data.
Furthermore, the blank-sky background files are recast in order to
have the same sky coordinates as RX J1347.  The background subtraction
(for spectra and surface brightness profiles) is performed as
described in full detail in \citet{arnaud02}.  This procedure consists
of two steps.  In a first step, for each product extracted from the
observation event list, an equivalent product is extracted from the
corresponding blank-field file and then subtracted from it.  This
allows us to remove the particle background.  However, if the
background in the observation region is different from the average
background in blank field data, this step could leave a residual
background component.  The residual background component is estimated
by using blank field subtracted data in a region free of cluster
emission and then subtracted in a second step from each MOS and pn
product.

The source and background events are corrected for vignetting using
the weighted method described in \citet{arnaud01}, the weight
coefficients being tabulated in the event list with the SAS task
\textit{evigweight}.  This allows us to use the on-axis response
matrices and effective areas.  Unless otherwise stated, the reported
errors are at 90\% confidence level.


\section{Surface brightness profile}
\label{morphology.sec}

Previous {\it Chandra} and {\it XMM-Newton} observations of RX J1347
revealed the presence of a hot and bright X-ray subclump visible to
the south-east (SE) of the main X-ray surface brightness peak
\citep{allen02b,gitti04}.  On the other hand, the data excluding the
SE quadrant (hereafter ``undisturbed cluster'') show a regular
morphology, indicating a relaxed state.  We are interested in
determining the characteristic properties of the cluster in order to
perform studies of mass profiles and scaling relations as it is
usually done for relaxed clusters. The disturbed SE quadrant is thus
masked in the following morphological analysis.
 
We compute a background-subtracted, vignetting-corrected, radial
surface brightness profile in the [0.3-2] keV energy band for each
camera separately. For the pn data, we generate a list of out-of-time
events\footnote{Out-of-time events are caused by photons which arrive
  while the CCD is being read out, and are visible in an uncorrected
  image as a bright streak smeared out in RAWY.}  (hereafter OoT) to
be treated as an additional background component. The effect of OoT in
the current observing mode (Full Frame) is 6.3\%. The OoT list is
processed in a similar way as done for the pn observation event
file. The profiles for the three detectors are then added into a
single profile, binned such that at least a signal-to-noise ratio of 3
is reached. The cluster emission is detected up to $R_{\rm out} =
1.73$ Mpc ($\sim 5$ arcmin).
\begin{figure}
\resizebox{\hsize}{!}{\includegraphics{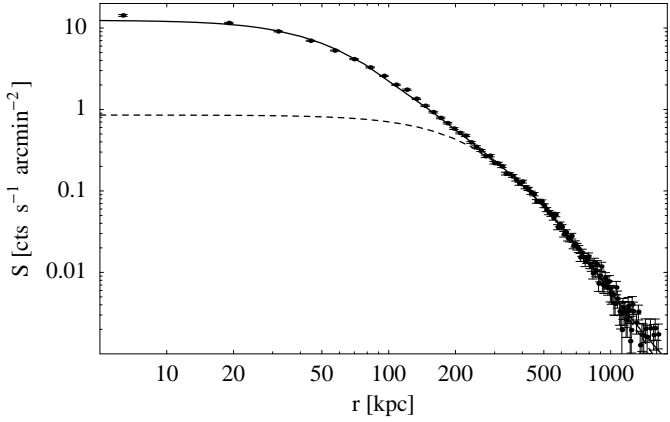}}
\caption{
    Background subtracted, azimuthally-averaged radial surface
    brightness profile in the [0.3-2] keV range of the data excluding
    the SE quadrant (undisturbed cluster).  The best fit $\beta$-model
    fitted over the $\sim$ 350 - 1730 kpc region is over-plotted as a
    dashed line (model SO in Table \ref{sxfits.tab}). When
    extrapolated to the center, this model shows a strong deficit as
    compared to the observed surface brightness. The solid line shows
    the best fit double $\beta$-model fitted over the whole region
    (model DD in Table \ref{sxfits.tab}).}
\label{Sx.fig}
\end{figure}
The surface brightness profile of the undisturbed cluster, shown in
Fig. \ref{Sx.fig}, is fitted in the CIAO tool \textit{Sherpa} with
various parametric models, which are convolved with the \textit{XMM}
point spread function (PSF).  The overall PSF is obtained by adding
the PSF of each camera \citep{ghizzardi01}, estimated at an energy of
1.5 keV and weighted by the respective cluster count rate in the
[0.3-2] keV energy band.  A single $\beta$-model \citep{cavaliere76}:
\begin{equation}
S(r) = S_0 \left( 1+ \frac{r^2}{r^2_{\rm c}} \right)^{-3\beta + 0.5}
\label{1beta-sb.eq} 
\end{equation}
is not a good description of the entire profile (model SG in Table
\ref{sxfits.tab}) and a fit to the outer regions (350 kpc \ltsim \, r
\ltsim \, 1730 kpc) shows a strong excess in the center as compared to
the model (see Fig. \ref{Sx.fig}).  The centrally peaked emission is a
strong indication of a cooling flow in this cluster.  We find that for
350 kpc \ltsim \, r \ltsim \, 1730 kpc the data can be described by a
$\beta$-model with a core radius $r_{\rm c}=307 \pm 9$ kpc and a slope
parameter $\beta=0.86 \pm 0.02$ (3$\sigma$ confidence level).  The
single $\beta$-model functional form is a convenient representation of
the gas density profile in the outer regions, which is used as a
tracer for the potential.  The parameters of this best fit are thus
used in the following to estimate the cluster gas and total mass
profiles in the region where the single $\beta$-model holds (see
Sect. \ref{mass.sec}).

We also consider a double isothermal $\beta$-model in the form:
\begin{equation}
S(r) = \sum_i S_{0,i} \left( 1+ \frac{r^2}{r^2_{{\rm c},i}} \right) 
       ^{-3\beta_i + 0.5}
\label{2beta-sb.eq} 
\end{equation}
where $i=1,2$, and find that it can account for the entire profile
(see Fig. \ref{Sx.fig}).
The best fit parameters are
$r_{\rm c,1}=39 \pm 1$ kpc, $\beta_1=0.62 \pm 0.01$,
$r_{\rm c,2}=386 \pm 17$ kpc, $\beta_2=1.01 \pm 0.05$.
By assuming a common $\beta$ value we find:
$r_{\rm c,1}=241 \pm 7$ kpc, $r_{\rm c,2}=47 \pm 2$ kpc,
$\beta=0.76 \pm 0.01$ (see Table \ref{sxfits.tab}).

\begin{table*}[!ht]
\caption{\label{sxfits.tab} 
Results from fitting the surface brightness profile of the undisturbed
cluster in different radial intervals [$R_{\rm in}$-$R_{\rm out}$].
The single and double $\beta$-models used for the fitting are given by 
Eqs. \ref{1beta-sb.eq} and \ref{2beta-sb.eq}, respectively.
  They are indicated with: 
  SG (\textit{S}ingle $\beta$-model, fitted in the \textbf{G}lobal radial range),
  SO (\textit{S}ingle $\beta$-model, fitted in the \textbf{O}uter region),
  DD (\textit{D}ouble $\beta$-model, with \textbf{D}ifferent $\beta$ values),
  DE (\textit{D}ouble $\beta$-model, with \textbf{E}qual $\beta$ values).
The quoted errors are at 3$\sigma$ confidence level.
}
\vskip 0.1truein
\hskip 0.0truein
\centerline{
\begin{tabular}{c|cc|cccc|c}
\hline
\hline
Model & \multicolumn{2}{|c|}{$R_{\mathrm{in}}$-$R_{\mathrm{out}}$} & $S_{\mathrm{0},i}$ & $\beta_i$ & \multicolumn{2}{c|}{$r_{\mathrm{c},i}$} & $\chi^2 / 
\mathrm{dof} \; (\chi^2_{\mathrm{red}})$
\\
 & (arcmin) & (kpc) & (cts/s/arcmin$^2$) & & (arcmin) & (kpc) & \\
\hline
\textbf{SG}: single $\beta$ & 0.0-5.0 & 0-1731 & $14.42^{+0.50}_{-0.50}$ & $0.590^{+0.005}_{-0.005}$ & $0.1492^{+0.0029}_{-0.0030}$ & $52^{+1}_{-1}$ & 1620/129 (12.56)
\\
\textbf{SO}: single $\beta$ & 1.0-5.0 & 346-1731 & $0.891^{+0.075}_{-0.075}$ & $0.861^{+0.022}_{-0.020}$ & $0.8876^{+0.0253}_{-0.0265}$ & $307^{+9}_{-9}$ & 109/87 (1.25)
\\
\textbf{DD}: double $\beta$ & 0.0-5.0 & 0-1731 & $18.94^{+0.86}_{-0.86}$ & $0.616^{+0.009}_{-0.008}$ & $0.1138^{+0.0032}_{-0.0033}$ & $40^{+1}_{-1}$ & 258/111 (2.32) 
\\
with $\beta_{\mathrm{1}} \neq \beta_{\mathrm{2}}$ & & & $0.42^{+0.04}_{-0.04}$ & $1.010^{+0.051}_{-0.043}$ & $1.1145^{+0.0483}_{-0.0506}$ & $386^{+17}_{-18}$ & 
\\
\textbf{DE}: double $\beta$ & 0.0-5.0 & 0-1731 & $18.12^{+1.00}_{-1.00}$ & & $0.1360^{+0.0048}_{-0.0050}$ & $47^{+2}_{-2}$ & 289/112 (2.58)
\\
with $\beta_{\mathrm{1}}=\beta_{\mathrm{2}}$ & & & $0.96^{+0.06}_{-0.06}$ & $0.761^{+0.012}_{-0.011}$ & $0.6968^{+0.0205}_{-0.0211}$ & $241^{+7}_{-7}$ &  
\\
\hline\noalign{\smallskip}  
\end{tabular}
}
\end{table*}


\section{Spectral analysis}
\label{spectral.sec}

Throughout the analysis, a single spectrum is extracted for each
region of interest and is then regrouped to reach a significance level
of at least 25 counts in each bin.  The data are modeled using the
XSPEC code, version 11.3.0 \citep{arnaud96}.  Unless otherwise
stated,the relative normalizations of the MOS and pn spectra are left
free when fitted simultaneously.  We use the following response
matrices: {\ttfamily m1{\_}439{\_}im{\_}pall{\_}v1.2.rmf} (MOS1),
{\ttfamily m2{\_}439{\_}im{\_}pall{\_}v1.2.rmf} (MOS2), {\ttfamily
  epn{\_}ff20{\_}sY9.rmf} (pn).


\subsection{Global spectrum}
\label{global.sec}

For each instrument, a global spectrum is extracted from all events
lying within 5 arcmin to the cluster emission peak.  We test in detail
the consistency between the three camera by fitting separately these
spectra with a {\ttfamily mekal} model (with the redshift fixed at
z=0.451) absorbed by a column density included in the {\ttfamily
  tbabs} model \citep[fixed at the nominal galactic value $N_{\rm H} =
4.85 \times 10^{20} \mbox{ cm}^{-2}$,][]{dickey90}.  Fitting the data
from all instruments above 0.3 keV leads to inconsistent values for
the temperature derived with the MOS and pn cameras: $kT =
12.2^{+0.7}_{-0.6}$ keV (MOS1), $10.4^{+0.5}_{-0.5}$ keV (MOS2),
$9.3^{+0.3}_{-0.3}$ keV (pn).  We then perform a systematic study of
the effect of imposing various high and low-energy cutoffs, for each
instrument.  Good agreement between the three cameras is found in the
[0.8-10.0] keV energy range ($kT = 11.2^{+0.7}_{-0.6}$ keV for MOS1,
$10.0^{+0.6}_{-0.5}$ for MOS2, $10.2^{+0.4}_{-0.4}$ for pn).  We
therefore perform the spectral analysis in this energy range.
The combined MOS+pn global temperature, in keV, and metallicity, as a
fraction of the solar value \citep{anders89} derived from the best fit
($\chi^2$/dof = 2717/1697) are respectively: $kT = 10.4^{+0.3}_{-0.3}$
keV, $Z = 0.25^{+0.03}_{-0.03} \, \rm{Z}_{\sun}$.  The unabsorbed
luminosities in this model (estimated from the average of the fluxes
measured by the three cameras after fixing $N_{\rm H}=0$) in the X-ray
([2.0-10.0] keV) and bolometric band are respectively: $L_X = 6.2 \pm
0.2 \times 10^{45} \mbox{ erg s}^{-1}$, $L_{\mbox{bol}}= 13.5 \pm 0.4
\times 10^{45} \mbox{ erg s}^{-1}$, where the errors are given as half
the difference between the maximum and the minimum value.


\subsection{Spatially resolved spectra}
\label{spectra.sec}

As done for the morphological analysis, for the spectral analysis we
separate the SE quadrant containing the X-ray subclump from the rest
of the cluster.  The data of the undisturbed cluster are divided into
the following annular regions: 0-30$''$, 30$''$-1$'$, 1$'$-1$'$.5,
1$'$.5-2$'$, 2$'$-3$'$, 3$'$-5$'$.  The spectra are modeled using a
simple, single-temperature model ({\ttfamily mekal} plasma emission
code in XSPEC) with the absorbing column density fixed at the nominal
Galactic value. The free parameters in this model are the temperature
$kT$, metallicity $Z$ \citep[measured relative to the solar values,
with the various elements assumed to be present in their solar
ratios,][]{anders89} and normalization (emission measure).  The
best-fitting parameter values and 90\% confidence levels derived from
the fits to the annular spectra are summarized in Table
\ref{profile_proj.tab}.

\begin{table}[ht]
\caption{
\label{profile_proj.tab}
Results of the spectral fitting in concentric annular
regions in the [0.8-10.0] keV energy range obtained by fixing the
absorbing column density to the Galactic value ($N_{\rm H} = 4.85
\times 10^{20} {\rm cm}^{-2} $). 
The temperature (in keV) and metallicity 
\citep[in fraction of the solar value,][]{anders89} are left as free parameters. 
The data of the SE quadrant are excluded (undisturbed cluster).
}
\vskip 0.1truein
\hskip 0.1truein
\begin{tabular}{c|cccc}
\hline
\hline
Radius & source counts & $kT$ & $Z$ & $\chi^2$/dof\\
(kpc) & (MOS+pn) & (keV) & ($Z_{\odot}$) & ~\\
\hline
0-173 & 46719 & $9.3^{+0.3}_{-0.3}$ & $0.31^{+0.05}_{-0.05}$
& 914/964
\\
173-346  & 18377 & $12.5^{+1.1}_{-0.9}$ &
$0.16^{+0.01}_{-0.01}$ & 573/546
\\
346-519 & 8733 & $11.8^{+1.5}_{-1.2}$ &
$0.22^{+0.14}_{-0.15}$ & 288/295
\\
519-692 & 4331 & $ 9.4^{+1.7}_{-1.3}$ &
$0.13^{+0.18}_{-0.13}$ & 201/178
\\
692-1038 & 4092 & $ 9.8^{+2.5}_{-1.7}$ &
$0.18^{+0.25}_{-0.18}$ & 315/229
\\
1038-1731 & 2742 & $ 7.3^{+4.2}_{-2.3}$ &
$0.40^{+0.64}_{-0.40}$ & 572/383 
\\
\hline 
\end{tabular}
\end{table}


\subsection{Deprojection analysis}
\label{deproj.sec}

Because of projection effects, the spectral properties at any point in
the cluster are the emission-weighted superposition of radiation
originating at all points along the line of sight through the
cluster. To correct for this effect, we perform a deprojection
analysis by adopting the XSPEC {\ttfamily projct} model. Under the
assumption of ellipsoidal (in our specific case, spherical) shells of
emission, this model calculates the geometric weighting factor
according to which the emission is redistributed amongst the projected
annuli.

The deprojection analysis is performed by fitting simultaneously the
spectra of the three cameras.  The results are reported in Table
\ref{deprojection.tab}.  We also calculate the electron density
$n_{\rm e}$ from the estimate of the Emission Integral $EI = \int
n_{\rm e} n_{\rm p} dV$ given by the {\ttfamily mekal} normalization:
$10^{-14} EI / ( 4 \pi [D_{\rm A} (1+z)]^2 )$. We assume $n_{\rm e} =
1.2023 n_{\rm p}$ in the ionized intra-cluster plasma.


\section{Radial profiles}
\label{radial.sec}


\subsection{Temperature}
\label{temp.sec}

\begin{figure}
\resizebox{\hsize}{!}{\includegraphics{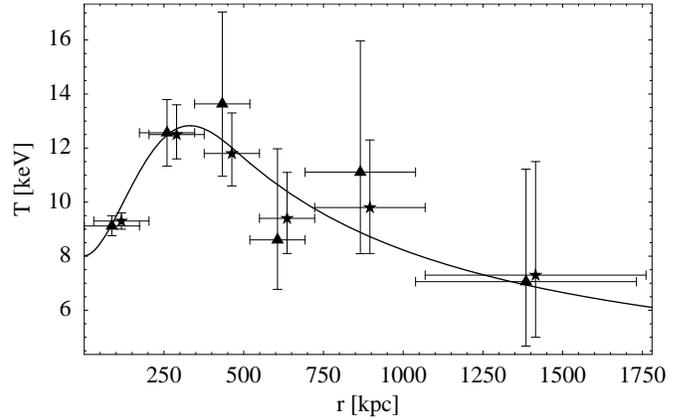}}
\caption{Deprojected (triangles) and projected (stars) X-ray gas
    temperature profiles measured in the [0.8-10.0] keV energy
    range. The data points of the projected profile are slightly
    shifted to the right to improve the clarity of the plot. The solid
    line shows the best fit function used in the total gravitational
    mass estimation presented in Sect. \ref{totmass.sec} below.}
\label{temp-cfr.fig}
\end{figure}

The deprojected temperature profile derived in Sect. \ref{deproj.sec}
is shown in Fig. \ref{temp-cfr.fig}, where we also show the projected
profile for comparison.  As expected, the deprojected central
temperature is lower than the projected one, since in the projected
fits the spectrum of the central annulus is contaminated by hotter
emission along the line of sight.  We also note that the projected
temperature profile measured by \textit{Chandra} \citep{allen02b} is
systematically slightly higher than that measured by
\textit{XMM-Newton}, although the general trend observed by the two
satellites is consistent \citep{gitti04}.

The temperature profile of RX J1347 exhibits the shape characteristic
for cool core clusters: the temperature declines from the maximum
cluster temperature at a break radius $r_{\mathrm{br}}$ moving
outwards and drops towards the cluster center.  If $r_{\mathrm{br}}$
is simply defined as the distance from the cluster center where the
temperature is maximal, then $r_{\mathrm{br}}=433 \pm 87$ kpc for the
deprojected profile and $r_{\mathrm{br}}=260 \pm 87$ kpc for the
projected profile, respectively.  This distance corresponds to $\sim
0.1-0.2 r_{\mathrm{vir}}$ (see Sect. \ref{rvir.sec}), in agreement
with works on the scaling properties of large samples of clusters of
galaxies \citep{markevitch,dm02,piff05,vikhlinin05,pratt2007}. The
temperature decrease observed in the outer regions ($\sim 40 \%$ from
$r_{\mathrm{br}}$ to $0.5 \, r_{\mathrm{vir}}$) is also consistent
with the findings of these studies.  The temperature derived from the
deprojected spectral analysis drops from the peak value of 13.6 keV to
the central minimum value of 9.1 keV.  This is fully consistent with
the typical 30\% drop seen in temperature profiles of cool core
clusters \citep[e.g. see][]{kaastra03}.

\begin{table}[ht]
\caption{
\label{deprojection.tab}
Results of the deprojection
analysis on annular MOS+pn spectra using the 
XSPEC {\ttfamily projct} model. The column density is
fixed to the Galactic value and the normalizations are in units of
$10^{-14} n_{\rm e} n_{\rm p} V / 4 \pi [D_{\rm A} (1+z)]^2$. The
fit gives $\chi^2$/dof = 3007/2557.
The data of the SE quadrant are excluded (undisturbed cluster).
}
\vskip 0.1truein
\hskip 0.1truein
\begin{tabular}{c|cccc}
\hline
\hline
Radius & $kT$ & $Z$ & norm & $n_{\rm e}$ \\
(kpc) & (keV) & ($Z_{\odot}$) & $(\times 10^{-3})$ &$(\times 10^{-3}$ cm$^{-3})$\\
\hline
0-173 & $9.1^{+0.4}_{-0.4}$ & $0.32^{+0.05}_{-0.05}$ 
& $6.02^{+0.08}_{-0.08}$ & $23.22^{+0.16}_{-0.16}$
\\
173-346 & $12.6^{+1.2}_{-1.2}$ &
$0.16^{+0.12}_{-0.15}$ & $2.73^{+0.09}_{-0.07}$ & $5.91^{+0.09}_{-0.08}$
\\
346-519 & $13.6^{+3.4}_{-2.7}$ &
$0.22^{+0.30}_{-0.21}$ & $1.51^{+0.07}_{-0.09}$ & $2.66^{+0.06}_{-0.08}$
\\
519-692 & $ 8.6^{+3.4}_{-1.8}$ &
$0.18^{+0.25}_{-0.18}$ & $0.88^{+0.06}_{-0.06}$ & $1.46^{+0.05}_{-0.05}$
\\
692-1038 & $ 11.1^{+4.9}_{-3.0}$ &
$0.08^{+0.39}_{-0.08}$ & $0.81^{+0.04}_{-0.07}$ & $0.69^{+0.02}_{-0.03}$
\\
1038-1731 & $ 7.1^{+4.2}_{-2.4}$ &
$0.39^{+0.59}_{-0.39}$ & $0.54^{+0.08}_{-0.07}$ & $0.25^{+0.02}_{-0.02}$
\\
\hline 
\end{tabular}
\end{table}


\subsection{Cooling time}
\label{tcool.sec}

\begin{figure}
\resizebox{\hsize}{!}{\includegraphics{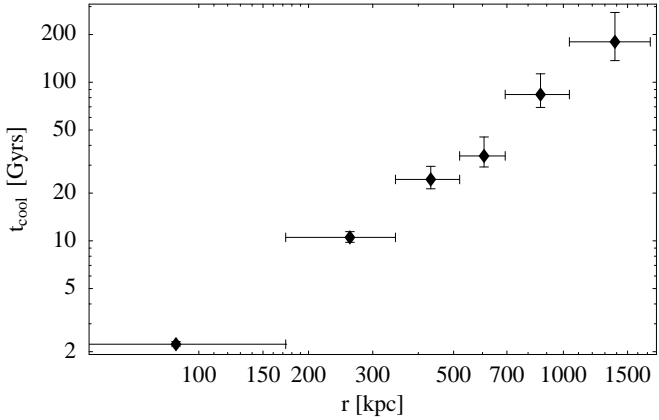}}
\caption{Cooling time as a function of radius.}
\label{fig:tcool}
\end{figure}

The cluster RX J1347 is known to host a cool core
\citep{schindler1997,allen02b,gitti04}. The centrally peaked surface
brightness profile and the central temperature drop discussed in
Sect. \ref{morphology.sec} and Sect. \ref{temp.sec}, respectively, are
indeed signatures of the presence of a central region where the plasma
cooling time is short. In the following we compute the cooling time
profile and the cooling radius of the cluster.

The cooling time is calculated as the characteristic time that it
takes a plasma to cool isobarically through an increment of
temperature $\delta T$:
\begin{equation}
t_{\rm cool} = \frac{5}{2} \frac{k \delta T}{n_{\rm e} \Lambda(T)} 
\label{tcool.eq}
\end{equation}
where $\Lambda(T)$ is the total emissivity of the plasma (the cooling
function) and $k$ is Boltzmann's constant.  Utilizing the deprojected
temperature profile and the density profile from Sect.
\ref{deproj.sec}, we can calculate the cooling time as a function of
radius, which is shown in Fig. \ref{fig:tcool}. The cooling time shows
a power law behavior as a function of radius. We find $t_{\rm cool}
\propto r^{1.46 \pm 0.01}$ when all 6 radial bins are used in the fit
and $t_{\rm cool} \propto r^{1.72 \pm 0.21}$ if only the 4 radial bins
beyond $0.2 \, r_{\mathrm{500}} \approx 280$ kpc are considered (see
Sect. \ref{rvir.sec} below for the definition and computation of
$r_{\mathrm{500}}$).  The latter value agrees with recent results from
the analysis in the same radial range of a sample of luminous clusters
at $z=0.2$ \citep{zhang07}. Following \cite{birzan}, we define the
cooling radius as the radius within which the gas has a cooling time
less than $7.7 \times 10^9$ yr, the look-back time to $z=1$ for our
adopted cosmology.  With this definition, we find $r_{\rm cool} \sim
210 \pm 10$ kpc which corresponds to the central 36 arcsec.

In the following analysis it is important to correct for the effects
of the central cooling flow when measuring the characteristic
temperature of the undisturbed cluster.  The average emission-weighted
cluster temperature is calculated by fitting with a {\ttfamily mekal}
model the spectrum extracted up to the outer radius detected by our
X-ray observation (5 arcmin), after excising the cooling region
(central 35 arcsec) and the SE quadrant.  We find a value $<T_{\rm X}>
= 10.1 \pm 0.7$ keV.


\subsection{Entropy}
\label{entropy.sec}

The gas entropy in groups and clusters of galaxies has recently
received particular attention since it resulted to be a very useful
quantity to probe the thermodynamic history of the hot gas in these
systems.  The entropy is usually defined as $S=k
T/n_{\mathrm{e}}^{2/3}$, where $T$ and $n_{\mathrm{e}}$ are the
deprojected electron temperature and density, respectively.

In cooling core clusters the radial entropy profiles are expected to
increase monotonically moving outwards, and to show no isentropic
cores \citep[e.g.,][]{mccarthy04}.  This behavior is indeed observed
in nearby cooling core clusters \citep{piff05,pratt06}.  Entropy
profiles are in general well described by a power law.  The value of
the power law index scatters around unity, depending on the cluster or
cluster sample used to derive it: for example, \cite{ettori02} found
0.97 from {\it Chandra} data of A1795, \cite{pratt04} derived a slope
of $0.94 \pm0.14$ from scalings of the entropy profiles of 5 clusters
observed with {\it XMM--Newton}, \cite{piff05} found $0.95 \pm0.02$
using scaled profiles of 13 cool core clusters observed with {\it
  XMM--Newton}, and \cite{pratt06} derived a slope of $1.08 \pm0.04$
(extending the sample studied in \cite{pratt04} to 10 objects).

\begin{figure}
\resizebox{\hsize}{!}{\includegraphics{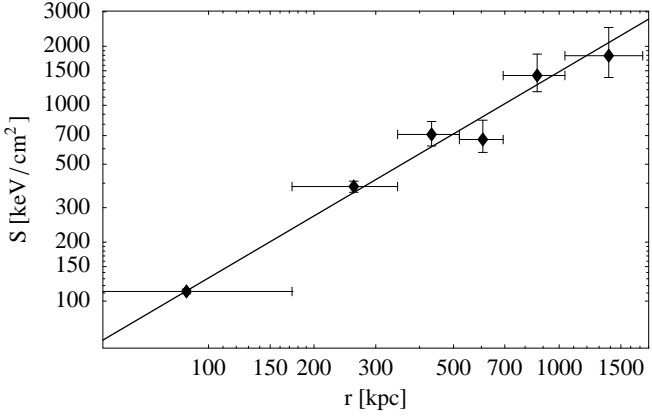}}
\caption{Entropy as a function of radius and the best fit power 
law $\mathrm{log}[S]=(1.053 \pm 0.005) \times \mathrm{log}[r] +
(0.011 \pm 0.010)$ (entropy in keV cm$^2$ and radius in kpc).}    
\label{fig:entropy}
\end{figure}

In Fig. \ref{fig:entropy} we show the gas entropy profile of RX J1347
computed from the deprojected temperature and electron density derived
in Sect. \ref{deproj.sec}.  We fit the profile with a line in log-log
space (with errors in both coordinates) and find:
$\mathrm{log}[S]=(1.053 \pm 0.005) \times \mathrm{log}[r] + (0.011 \pm
0.010)$ (entropy in keV cm$^2$ and radius in kpc), which is consistent
with previous results. \cite{megan06} recently found that the entropy
profiles they derived from {\it Chandra} observations of 9 cool core
clusters are better fitted by a power law plus a constant entropy
pedestal of $\approx 10$ keV cm$^2$ than by a pure power law. We
performed similar fits and find an entropy pedestal consistent with
zero. However, we notice that this result might be due to the lack of
adequate spatial resolution of the entropy profile in the central
region.

Recent results suggest that the entropy scales with the temperature as
$S \propto <T_{\rm X}>^{0.65}$, the so-called ``entropy ramp'',
instead of the self-similar scaling $S \propto <T_{\rm X}>$
\citep{ponman03,pratt04,piff05,pratt06}. Here $<T_{\rm X}>$ is the
mean cluster/group temperature corrected for the cool core effect and
$S$ is the entropy measured at some fraction of he virial radius
(usually $0.1 \times r_{\mathrm{200}}$, see Sect. \ref{rvir.sec} below
for the definition and computation of $r_{\mathrm{200}}$). In order to
verify if the entropy measured in RX J1347 follows this relation, we
therefore adopt the scaling $S \propto h^{-4/3}(z) \, (<T_{\rm X}>/10
\, \mathrm{keV})^{0.65}$, with a mean temperature for RX J1347 equal
to $10.1 \pm 0.7$ keV (see Sect. \ref{tcool.sec}).  Here
$h^2(z)=\Omega_{\mathrm{m}} (1+z)^3+\Omega_{\mathrm{\Lambda}}$ and the
factor $h^{-4/3}$ comes from the scaling of the density. At $0.1
\times r_{\mathrm{200}}$ the scaled entropy is equal to $382 \pm 32$,
$349 \pm 54$, and $437 \pm 51$ keV cm$^2$ for $r_{\mathrm{200}}$
computed from the total mass profiles derived from model SO, DDg1, and
NFW, respectively (see Sect. \ref{totmass.sec} below for the different
models used in the total mass determination from the X-ray data). If
instead the value $r_{\mathrm{200}}^{\mathrm{Sim}}$ is used (i.e., we
adopt the size-temperature relation calibrated through numerical
simulations, see Sect. \ref{rvir.sec} below), the normalization is
$567 \pm 70$ keV cm$^2$. The errors on these normalizations also take
into account the uncertainty in the estimate of
$r_{\mathrm{200}}$. The normalization derived by adopting the
size-temperature relation is in good agreement with the entropy
normalization of the $S(0.1 \times r_{\mathrm{200}})$-$<T_{\rm X}>$
relation at $<T_{\rm X}>=10 \, \mathrm{keV}$ by \citep[][see their
Fig. 4]{ponman03}. The values computed using $r_{\mathrm{200}}$
derived from the total mass profiles are smaller, but still consistent
within the uncertainties, than the values found by \cite{ponman03}.

\section{Mass determination}
\label{mass.sec}

In \cite{gitti04} we presented the total mass profile estimated from
the single $\beta$-model.  Here we perform a detailed study of the
radial profiles of total gravitational mass and gas mass reconstructed
by using different methods.  The new values do not change the main
conclusions in \cite{gitti04} but are more accurate.  In this section
we also present the computation of the characteristic radii
$r_{\mathrm{\Delta}}$ quoted above.


\subsection{Total gravitational mass}
\label{totmass.sec}

The analysis to estimate the total gravitational mass of RX J1347 is
not limited to only one specific method, but is instead carried out by
adopting different approaches. This enables us to investigate the
effects introduced by different fitting functions for the gas density
and temperature, and different methods to derive the total mass from
the observed gas distribution.

The total gravitating mass distribution is calculated under the usual
assumptions of hydrostatic equilibrium and spherical symmetry by using
\begin{equation}
M_{\mathrm{tot}}(<r)= - \frac{kT(r) \, r}{G \mu m_p} 
\left[ \frac{ d \ln \rho_g(r)}{d \ln r} +
\frac{d \ln T_g(r)}{d \ln r} \right]
\label{mass.eq}
\end{equation}
where $G$ and $m_p$ are the gravitational constant and proton mass and 
$\mu = 0.62$.
A welcome property of Eq. \ref{mass.eq} is that the total
gravitational mass within a sphere of radius $r$ is determined from
the gas density $\rho_{\mathrm{g}}$ and temperature $T_{\mathrm{g}}$
measured at the cluster-centric distance $r$.  This implies that when
the gas density and temperature are well modeled only in the radial
range $R_{\mathrm{in}}-R_{\mathrm{out}}$ but not within
$R_{\mathrm{in}}$, the mass determination is still reliable in the
range $R_{\mathrm{in}}-R_{\mathrm{out}}$.  As shown in
Sect. \ref{morphology.sec}, a single $\beta$-model provides a good fit
to the surface brightness profile in the radial range 350 kpc \ltsim
\, r \ltsim \, 1730 kpc (model SO in Table \ref{sxfits.tab}).  In this
case the deprojected gas density profile is easily computed and the
total cluster mass is independent of the gas density central
value. Since beyond 350 kpc the temperature profile is declining, it
can be well modeled through the polytropic relation $T \propto
\rho_{\mathrm{g}}^{\gamma - 1}$, with $1 \leq \gamma \leq 5/3$.  The
polytropic fit to the deprojected temperature profiles gives in this
case $\gamma=1.23 \pm 0.02$ ($1\sigma$ error on one parameter).  The
total mass profile computed using this model is discussed below
together with the results from the more sophisticated double
$\beta$-model.

In order to obtain a total mass estimate for the whole observed radial
range we use the double $\beta$-model fits discussed in Sect.
\ref{morphology.sec} (model DD and DE in Table \ref{sxfits.tab}).  The
gas density is computed from the double $\beta$-model surface
brightness fits using the formulas derived in \citet{xuewu00}: we
assume that each component corresponds to a gas phase, invert
Eq. \ref{2beta-sb.eq} and compute the the electron number densities
for the two components $n_{\mathrm{e,i}}(r)$ and the total electron
number density, $n_e(r)$ using:
\begin{equation}
n_e(r) = \sum_i n_{e,i}(r) = \left[ n_e(0) \sum_i \tilde{n}_{e,i}(r) 
          \right]^{1/2}, 
\end{equation}
\begin{equation}
n_{e,i}(r) = \left[ \frac{n_e(0)}{n_e(r)} \right] ~\tilde{n}_{e,i} (r),
\end{equation}
\begin{equation}
\tilde{n}_{e,i}(r) = n_{e,i}(0) 
\left( 1 + {r^2 \over r_{c,i}^2} \right)^{-3\beta_i},
\end{equation}
where i=1,2 and $n_e(0)$ is the central, total electron density.
The central number densities for the two components are given by
\begin{equation}
n^2_{e,i}(0) = \left[ \frac{4\pi^{1/2}}{\alpha(T_i) g_i \mu_e} \right]
              ~\left[ \frac{\Gamma(3\beta_i)}{\Gamma(3\beta_i - 1/2)}
               \right]
              ~\left( \frac{S_{0,i}}{r_{c,i}} \right) ~A_{ij}
\label{eqn:n0}
\end{equation}
in which
\begin{equation}
\frac{1}{A_{ij}}   = 1 + \left( \frac{g_i}{g_j} \right)
                   \left( \frac{r_{c,i} S_{0,j}}{r_{c,j} S_{0,i}} \right)
                   \left( \frac{T_i}{T_j} \right)^{1/2}
                  ~\left[ \frac{\Gamma(3\beta_j) ~\Gamma(3\beta_i - 1/2)}
                    {\Gamma(3\beta_i) ~\Gamma(3\beta_j - 1/2)} 
                   \right] , 
\label{eqn:Aij}
\end{equation}
where j=1,2 and $j\not=i$.  Here $g_i$ is the Gaunt factor for the
component $i$ and $\alpha(T_i)$ is the emissivity due to thermal
bremsstrahlung.  The Gaunt factors are computed using the results of
\citet{sutherland}.  Note that in the derivation of the equations
given above it is assumed that each component has a constant electron
temperature $T_i$ throughout the cluster.  As shown in
Sect. \ref{radial.sec} the gas is not isothermal hence this assumption
is not strictly valid.  Nevertheless the temperature dependence of the
above equation is fairly weak and we set $T_{1 \mathrm{or}
  2}=T_{\mathrm{max}}=13.6 \, \mathrm{keV}$ (the maximum of the
temperature profile) and $=T_{2 \mathrm{or} 1}=T_{\mathrm{min}}=7.1 \,
\mathrm{keV}$ (the minimum of the temperature profile) to quantify the
maximum variation of the total mass estimate with temperature.  Using
the above equations and Eq. \ref{mass.eq} we compute the mass profile
for 4 cases: DDg1 (model DD and $T_{1}=T_{\mathrm{max}},
T_{2}=T_{\mathrm{min}}$), DDg2 (model DD and $T_{1}=T_{\mathrm{min}},
T_{2}=T_{\mathrm{max}}$), DEg1 (model DE and $T_{1}=T_{\mathrm{max}},
T_{2}=T_{\mathrm{min}}$), and DEg2 (model DE and
$T_{1}=T_{\mathrm{min}}, T_{2}=T_{\mathrm{max}}$).  While the
assumption of isothermality is justified in the evaluation of the
density-dependent term of Eq. \ref{mass.eq} from the observed surface
brightness profile, the radial dependence of the gas temperature must
be carefully modeled, since the total gravitational mass varies
strongly with temperature.  The temperature profile in the whole
observed range is clearly not well described by a polytropic relation and it
is not possible to model it using a single analytical function due to
the central temperature drop.  We therefore model the profile using
two functions joined smoothly at a cut radius $R_{\mathrm{cut}}$,
i.e. we take care that the temperature profile and its gradient are
continuous across $R_{\mathrm{cut}}$.  Since the polytropic relation
provides a good description in the outer region, we adopt $T \propto
\rho_{\mathrm{g}}^{\gamma - 1}$ as fitting function for $r \geq
R_{\mathrm{cut}}$, with $\rho$ computed from the double $\beta$-model
fits.  The values obtained for the parameter $\gamma$ are very similar
to those obtained when using the single $\beta$-model.  Within
$R_{\mathrm{cut}}$ we choose to fit the temperature profile using a
5-th order polynomial with zero derivative at the center.  If the
latter condition is not satisfied the derived total mass density is
found to be negative in the cluster core.  We vary $R_{\mathrm{cut}}$
and find that $R_{\mathrm{cut}}=520 \, \mathrm{kpc}$ provides the best
model. The resulting best fit function is shown in
Fig. \ref{temp-cfr.fig}. The total mass profiles computed from the
surface brightness fits presented in the following are computed using
this temperature profile modeling and will be indicated by the name of
the model used to describe the surface brightness (see Table
\ref{sxfits.tab}).

\begin{figure}
\vskip -0.8cm
\resizebox{\hsize}{!}{\includegraphics{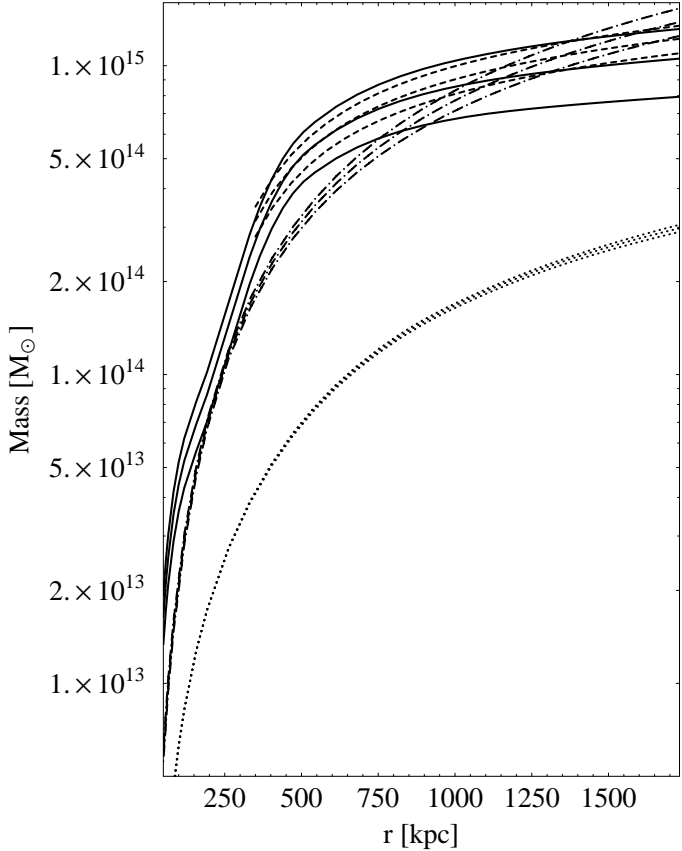}}
\vskip -1cm
\caption{Integrated three dimensional total mass profiles, with
    errors, derived from the double $\beta$-model (model DDg1, solid),
    single $\beta$-model (model SO, dashed), and the NFW model
    (dot-dashed). The dotted line shows the cumulative gas mass
    profile. See text for details.}
\label{fig:mtot}
\end{figure}

The relative difference between the mass profiles for model DDg1 and
DDg2 (DEg1 and DEg2) is less than 4 $\%$ (6 $\%$) in the whole
observed radial range (0-1731 kpc).  Models DDg1 and DDg2, and DEg1
and DEg2 give nearly identical results for $r > 500$ kpc.  The largest
difference is found between models DDg1 and DEg2, but it is less that
15 $\%$ in the whole radial range and less than 5 $\%$ for $r > 250$.
These small differences show that the temperature does not
significantly affect the gas density determination for this massive
and hot cluster, and that models DE an DD provide the same mass
estimate for the whole radial range of interest.  Given these results
and the fact that model DD gives a smaller $\chi^2_{\mathrm{red}}$
than model DE for the surface brightness modeling, we will discuss, in
the following, only the mass profile derived using model DDg1.  We
compare the mass profiles derived from the double $\beta$-model with
the one from the single $\beta$-model in the radial range 350-1731
kpc.  In this range the relative difference of the mass profiles is at
most 13 $\%$ (close to the innermost and outermost radii), but smaller
than 10 $\%$ in the range 380-1500 kpc for the four double
$\beta$-models we derived.  Hence, the double $\beta$-model provides
estimates in good agreement with the single $\beta$-model, and is of
course preferred since it allows us to estimate the mass in the whole
observed radial range, i.e. 0-1731 kpc.  The mass profiles from the
double $\beta$-model (model DDg1) and the single $\beta$-model (model
SO) are plotted in Fig. \ref{fig:mtot}.  Errors on the total
gravitational masses are computed by propagating the $1 \sigma$ errors
on the surface brightness and temperature profiles best fit
parameters, and are of the order of 10 $\%$ and 20 $\%$ for the values
derived from the single and double $\beta$-model, respectively.  The
profile derived using the single $\beta$-model is shown only in the
region where it is valid, i.e. for $r>350$.  The mild depression
visible around $\sim 250$ kpc in the mass profile derived from the
double $\beta$-model is due to the shape of the temperature profile in
the inner region.

The cluster gravitational mass can also be computed by making direct
use of the gas temperature and gas density profiles derived from the
deprojection analysis presented in Sect.~\ref{deproj.sec}.  We invert
the equation of hydrostatic equilibrium (Eq.\ref{mass.eq}) and, using
the three-dimensional gas density, we select the dark matter mass
model that reproduces better the deprojected temperature profile.  In
the minimization the $1 \sigma$ errors on one single parameter from
the spectral fits are used.  For dark matter mass model, we consider
the integrated NFW \citep{nfw96} dark matter profile:
\begin{equation}
\label{NFW}
M_{\mathrm{DM}}(<r)=4 \pi r_{\mathrm{s}}^3 \rho_{\mathrm{c,z}} \frac{200}{3} 
\frac{c^3\Big( \mathrm{ln}(1+r/r_{\mathrm{s}}) -\frac{r/r_{\mathrm{s}}}
{(1+r/r_{\mathrm{s}})}\Big)}{\mathrm{ln}(1+c)-c/(1+c)},
\end{equation}
where $ \rho_{{\rm c},z} = (3 H_z^2)/ (8 \pi G)$ is the critical
density at the cluster's redshift.  The scale radius $r_{\mathrm{s}}$
and the concentration parameter $c$ are the free parameters.  The
total gravitational mass within a sphere of radius $r$ is given by gas
plus dark matter mass and therefore
$M_{\mathrm{tot}}(<r)=M_{\mathrm{gas}}(<r)+M_{\mathrm{DM}}(<r)$ in
Eq.(\ref{mass.eq}).  Nevertheless, in most of the work
$M_{\mathrm{tot}}(<r)=M_{\mathrm{DM}}(<r)$ is used, i.e. the NFW
profile is used to fit dark matter plus gas mass.  We also computed
the total mass profile by taking into account the gas mass, i.e. by
adding the cumulative gas mass profile to the best-fitting NFW
profile, and found little difference between the two profiles.  The
best-fit parameters are $r_{\mathrm{s}}=722 \pm 112$ kpc and $c=3.20
\pm 0.30$ (errors are RMS of the 1$\sigma$ joint confidence limits),
with $\chi^2_{\mathrm{min}}=6.7$ for 4 degrees of freedom.  Our
best-fit NFW profile is shown in Fig. \ref{fig:mtot}. From the set of
$(c,r_{\mathrm{s}})$ parameters acceptable at 1 $\sigma$ we compute,
for each radius, the maximum and minimum value of the total mass and
hence its upper and lower errors.  These are of the order of 10
$\%$. From a visual inspection of Fig. \ref{fig:mtot} one can note
that the NFW mass profile is lower than the double $\beta$ estimate
for $r<1150$ kpc and higher at larger radii.  The discrepancy within
1150 kpc is due to the fact that our best-fit NFW profile tends to
underestimate the temperature in this range.  The relative difference
between the NFW and the double $\beta$ mass profiles is -38 $\%$
(underestimate) at $r=500$ but decreasing towards the center, and
increases almost linearly to 30 $\%$ (overestimate) at $r=1731$ kpc.
The fairly low concentration parameter $c$, compared to the
predictions of numerical simulations \citep[e.g,][]{maccio2006}, and
the goodness of our NFW fit might indicate that our temperature
profile is not enough spatially resolved in the central region of the
cluster for this kind of mass determination method.  While the mass
determination from the double $\beta$-model may therefore be
preferred, we present values also from the NFW fitting for
completeness.

\subsection{Virial radius and scaling relations}
\label{rvir.sec}

In this section we determine the characteristic radii
$r_{\mathrm{\Delta}}$ used in Sects. \ref{radial.sec} and
\ref{entropy.sec}.  For the various mass profiles we compute
$r_{\mathrm{\Delta}}$, the radius within which the mean interior
density is $\Delta$ times the critical value, by using
\begin{equation}
\label{rdelta}
\Delta=\frac{3 M_{\mathrm{tot}}(<r_{\Delta})}{4 \pi \rho_{c,z}r_{\Delta}^3 }.
\end{equation} 
For the cosmology adopted here the virial radius is given by
$r_{\mathrm{vir}}=r_{\mathrm{\tilde{\Delta}}}$, with
$\mathrm{\tilde{\Delta}}=178+82 x -39 x^2$ and where $x=\Omega(z)-1$
and $\Omega(z)=0.3 \, (1+z)^3/(0.3 \, (1+z)^3+0.7)$
\citep{bryanandnorman}.  Thus for RX J1347
$\mathrm{\tilde{\Delta}}=135$.  We also compute
$M_{\rm{tot}}(<r_{\Delta})$ and $M_{\rm{gas}}(<r_{\Delta})$ for
various overdensities: $\Delta = 2500, 1000, 500, 200$.  The results
obtained from the overdensity profiles calculated from the double
$\beta$-model (DDg1) and NFW fit are reported in Table
\ref{delta.tab}.

\begin{table*}[!ht]
\caption{\label{delta.tab} 
Characteristic radii $r_{\Delta}$, total mass $M_{\rm tot}$ and gas mass $M_{\rm gas}$ 
for various overdensities $\Delta$ derived from the double $\beta$-model (DDg1) 
and NFW fits (1$\sigma$ errors in parentheses). 
The masses are estimated within $r_{\Delta}$.
As discussed in Sect. \ref{totmass.sec}, results from the double $\beta$-model 
are generally more reliable. 
}
\vskip 0.1truein
\hskip 0.0truein
\centerline{
\begin{tabular}{c|ccc|ccc}
\hline
\hline
$\Delta$ & $r_{\rm \Delta,DDg1}$ & $M_{\rm tot,DDg1}$ & $M_{\rm gas,DDg1}$ & $r_{\rm \Delta,NFW}$ & $M_{\rm tot,NFW}$ & $M_{\rm gas,NFW}$
\\
~ & (kpc)    & ($10^{14} M_{\odot}$) &  ($10^{14} M_{\odot}$) &  (kpc) & ($10^{14} M_{\odot}$) &  ($10^{14} M_{\odot}$) 
\\
\hline
200 & 1957.2 (183.2) & 11.00 (2.78) & 3.34 (0.11) & 2286.7 (110.8) & 17.86 (2.07) & 3.82 (0.18) 
\\
500 &  1387.2 (123.9) & 9.77 (2.30) & 2.39 (0.05) & 1479.2 (71.23) & 11.85 (1.13) & 2.55 (0.06) 
\\
1000 & 1063.3 (91.00) & 8.80 (1.91) & 1.79 (0.03) & 1029.3 (43.08) & 7.99 (0.62) & 1.73 (0.03)
\\
2500 & 729.3 (63.2) & 7.10 (1.40) & 1.15 (0.02) & 608.1 (19.8) & 4.12 (0.22) & 0.91 (0.01)
\\ 
\hline\noalign{\smallskip}  
\end{tabular}
}
\end{table*}

The size-temperature relation $r_{\mathrm{\Delta}} \propto
\sqrt{<T_{\rm X}>}$ predicted by self-similarity allows an estimate of
$r_{\mathrm{\Delta}}$ from the mean cluster temperature alone,
provided that its normalization is known from numerical simulations.
We compute the normalization for the cosmology adopted here by
interpolating the values given in \citet{evrard96}.  For the mean
cluster temperature $<T_{\rm X}>=10.1 \pm 0.7$ keV we derive the
characteristic radii $r_{\mathrm{\Delta}}^{\mathrm{Sim}}$, finding
$r_{\mathrm{2500}}^{\mathrm{Sim}}= 886 \pm 30$ kpc and
$r_{\mathrm{vir}}^{\mathrm{Sim}}= 3197 \pm 107$ kpc.  From our X-ray
analysis we find $r_{\mathrm{2500}}=(734 \pm 34, 729 \pm 63, 608 \pm
20)$ kpc and $r_{\mathrm{vir}}=(2378 \pm 76, 2241 \pm 189, 2639 \pm
108)$ kpc when using in Eq. \ref{rdelta} the mass profile derived from
model (SO, DDg1, NFW), respectively.  These values are consistent with
the size-temperature relation derived from observations of nearby
relaxed clusters \citep{arnaud05}.
By comparing the above values we note that the estimates from the
X-ray analysis are systematically lower than the ones predicted from
the size-temperature relation calibrated by means of numerical
simulations.  It is not surprising that we find a smaller discrepancy
for $r_{\rm 2500}$ than $r_{\rm vir}$, as its determination does not
require extrapolation of the observed mass profile.  This is in
agreement with results for other individual clusters
\citep[e.g.,][]{gitti07} and studies of cluster samples
\citep[][]{sanderson03,piff05}.  The largest discrepancy is found for
$r_{\mathrm{vir}}$ and in poor, cool clusters.  In these systems the
impact of additional, non-gravitational heating is most pronounced, as
the extra energy required to account for their observed properties is
comparable to their thermal energy \citep[][]{ponman96,tozzi01}.
The observed discrepancy is also related to the cluster total mass
determination.  In this context it is interesting to note that recent
results from numerical simulations indicate that the total mass of
simulated clusters estimated through the X-ray approach is lower that
the true one due to gas bulk motions (i.e. deviation from the
hydrostatic equilibrium) and the complex thermal structure of the gas
\citep{rasia2006,nagai2007}.

A self-similar scaling relation between $M_{\rm tot}$ and $<T_{\rm
  X}>$ at a given overdensity is predicted in the form $M_{\rm tot}
\propto <T_{\rm X}>^{3/2}$.  Various observational studies have found
different and sometime conflicting results regarding the slope and
normalization of the $M$-$T$ relation \citep[e.g.,][and references
therein]{allen01,finoguenov01,ettori02b,sanderson03, arnaud05}.  The
relation derived by \citet{arnaud05} for a sub-sample of six relaxed
clusters hotter than 3.5 keV observed with \textit{XMM-Newton} is
consistent with the standard self-similar expectation, following the
relation:
\begin{equation}
h(z) M_{2500} = (1.79 \pm 0.06)\times 10^{14} {\rm M}_{\odot} 
\left( \frac{<T_{\rm X}>}{5 \, {\rm keV}} \right)^{1.51 \pm 0.11} 
\label{m2500.eq}
\end{equation}
This result is in agreement with \textit{Chandra} observations
\citep{allen01}.  In the case of RX J1347, Eq. \ref{m2500.eq} turns
into an estimate of $M_{\rm 2500} = (4.07 \pm 0.46) \times 10^{14}
M_{\odot}$.  By considering the whole \textit{XMM-Newton} sample (ten
clusters in the temperature range [2-9] keV), the relation steepens
with a slope $\sim 1.70$ \citep{arnaud05} indicating a breaking of
self-similarity.  In this case we estimate $M_{\rm 2500, DDg1} = (4.39
\pm 0.35) \times 10^{14} M_{\odot}$.  The mass estimate that we derive
at the overdensity $\Delta = 2500$ differs strongly depending on the
model adopted (see Table \ref{delta.tab}).  From model DDg1 we
estimate $M_{\rm 2500,DDg1} = (7.10 \pm 1.40) \times 10^{14} M_{\odot}$,
which is much higher than the prediction of the $M$-$T$ relation.  The
mass estimate of $M_{\rm 2500, NFW} = (4.12 \pm 0.22) \times 10^{14}
M_{\odot}$ as derived from the best-fitting NFW profile is instead in
good agreement with the $M$-$T$ relation, although the large error
bars prevent us from distinguishing between a self-similar or steeper
relation.


\subsection{Gas mass and gas mass fraction}
\label{gasmass.sec}

From the results of the deprojected spectral analysis we compute the
cumulative gas mass profile $M_{\mathrm{gas}}(<r)$, thus obtaining
values for the 6 bins used in in the spectral analysis.  In order to
derive better estimates when an extrapolation of the gas mass beyond
$R_{\mathrm{out}}$ is needed, we compute the gas mass profile using
the radial gas density profile derived from the best fit parameters of
the double $\beta$-model (model DDg1) of the surface brightness
profile.  The normalization of the latter is fixed using the gas
density profile from the spectral analysis.  The resulting gas mass
profile is shown in Fig. \ref{fig:mtot}.  When $M_{\mathrm{gas}}(<r)$
is evaluated within $R_{\mathrm{out}}$ we use the binned profile and
spline interpolation, which in this radial range provides values
consistent with the ones computed using the results from the double
$\beta$-model.

The gas mass fraction $f_{\rm gas}$ is defined as the ratio of the
total gas mass to the total gravitating mass within a fixed volume.
We measure $f_{\rm gas, 2500} = 0.162 \pm 0.036$ from the mass
profiles derived from the double $\beta$-model fit (model DDg1).  This
value is close to the global baryon fraction in the Universe,
constrained by CMB observations to be $\Omega_b / \Omega_m = 0.175 \pm
0.023$ \citep{readhead04,spergel03}, and is higher than the average
value derived in a number of previous measurements with
\textit{Chandra} \citep[e.g.,][]{allen02a,vikhlinin06}.  However, we
note that a general trend of increasing $f_{\rm gas}$ with cluster
temperature (hence mass) has been observed \citep{vikhlinin06}.  The
high central gas mass fraction measured here is consistent with this
tendency, as RX J1347 is a hot, massive cluster.


\section{Comparison with previous work}
\label{comp.sec}

In this section we compare the most relevant total and gas mass
estimates for RX J1347 found in literature with our results.  The
values in literature are converted to the cosmology adopted here
before the comparison.

\subsection{Comparison with X-ray studies}

Using combined ROSAT and ASCA observations \citet{schindler1997}
derived $M_{\mathrm{tot}}=1.11 \times 10^{14} M_{\sun}$,
$M_{\mathrm{tot}}=4.93 \times 10^{14} M_{\sun}$, and
$M_{\mathrm{tot}}=1.45 \times 10^{15} M_{\sun}$ within 204, 850, 2550
kpc, respectively.  
These values were derived assuming isothermality
and the error coming from the uncertainty on the global temperature is
of the order of $10\%$-$15\%$,
as we estimated from the plot showing the profile of the integrated total
mass \citep[see][Fig. 6]{schindler1997}.
We find, for model (SO, DDg1), $M_{\mathrm{tot}}=(1.14 \pm 0.14, 0.93
\pm 0.17) \times 10^{14} M_{\sun}$, $M_{\mathrm{tot}}=(8.10 \pm 0.84,
7.85 \pm 1.60) \times 10^{14} M_{\sun}$, and $M_{\mathrm{tot}}=(1.47
\pm 0.15, 1.22 \pm 0.32)\times 10^{15} M_{\sun}$ within 204, 850, 2550
kpc, respectively. While we find a significant mismatch at 850 kpc,
the results are in reasonably good agreement at small and large radii,
in particular considering the errors and the different assumptions
adopted in the mass determination. For the cumulative gas mass
\citet{schindler1997} found $M_{\mathrm{gas}}=1.33 \times 10^{14}
M_{\sun}$ and $M_{\mathrm{gas}}=5.93 \times 10^{14} M_{\sun}$ within
850, 2550 kpc respectively, while our values are
$M_{\mathrm{gas}}=(1.39 \pm 0.02) \times 10^{14} M_{\sun}$ and
$M_{\mathrm{gas}}=(4.32 \pm 0.17) \times 10^{14} M_{\sun}$ within 850,
2550 kpc, respectively. While the values at 850 kpc are consistent,
the large value found at 2550 by \citet{schindler1997} is very likely
due to the narrower radial range probed by their observation. As shown
in Sect. \ref{morphology.sec} (see Table \ref{sxfits.tab}), the gas
density steepens in the outer region. As a result, the gas mass
derived from a single $\beta$-model fit to a narrow central region and
extrapolated to large radii is biased high. In comparing the results,
we should also bear in mind that the analysis presented by
\citet{schindler1997} is performed on the full $360^{\circ}$ data, as
the hot enhancement in the SE quadrant has been discovered only
subsequently with \textit{Chandra} and \textit{XMM-Newton}
observations \citep{allen02b,gitti04}.

We compare our best-fitting NFW profile with the one derived by
\citet{allen02b} from {\it Chandra} data.  The two profiles are
consistent, with a relative difference ranging from 15$\%$ to 30$\%$
depending on the radial range considered.  As a general trend, our
profile results lower in the inner region (inside $\sim$600 kpc) and
higher in the outer region (outside $\sim$1000 kpc) than the one
derived by \citet{allen02b}.  In particular, \citet{allen02b} find an
integrated mass within the virial radius of their best-fitting NFW
mass profile of $M_{\mathrm{tot}}(<2
\mathrm{Mpc})=(1.95^{+1.48}_{-0.70})\times 10^{15} M_{\sun}$, which is
in fairly good agreement with the value that we measure:
$M_{\mathrm{tot}}(<2 \mathrm{Mpc})=(1.59^{+0.18}_{-0.16})\times
10^{15} M_{\sun}$.  When considering the mass profile derived from the
double $\beta$-model (DDg1), which at large distances is lower than
the one derived from the NFW fit (see Fig. \ref{fig:mtot} ), we find
$M_{\mathrm{tot}}(<2 \mathrm{Mpc})=(1.11^{+0.28}_{-0.27})\times
10^{15} M_{\sun}$.  This value is fairly low compared to the value
found by \citet{allen02b}, but still consistent considering the errors
on the mass estimates at this large distance.

\citet{ettori2004} derive from {\it Chandra} data estimates of
$M_{\mathrm{tot}}=(8.94 \pm 0.80) \times 10^{14} M_{\sun}$ and
$M_{\mathrm{gas}}=(1.81 \pm 0.08) \times 10^{14} M_{\sun}$ within 1368
kpc, which corresponds to $r_{\mathrm{500}}$ in their work.  While our
value $M_{\mathrm{tot}}(<1368 \rm{ kpc})=(10.84 \pm 1.11, 9.72 \pm
2.27) \times 10^{14} M_{\sun}$ (for model SO and DDg1, respectively)
agrees with the {\it Chandra} estimate, we find a larger value for the
gas mass: $M_{\mathrm{gas}}(<1368 \rm{ kpc})=(2.35 \pm 0.05) \times
10^{14} M_{\sun}$.  The discrepancy might be related to the different
approaches adopted for the calculation.  We estimate the gas mass
directly from the density profile derived from the deprojected
spectral analysis (Sect. \ref{deproj.sec}).  The gas mass computed by
\citet{ettori2004} is derived by estimating the central electron
density from the combination of the best-fit results of the spectral
and imaging analyses (namely the normalization of the thermal spectrum
and the parameters of the single $\beta$-model).  In particular, the
low value measured by \citet{ettori2004} might be biased by an
underestimate of the central density due to a possible undersampling
of the cluster luminosity within the radius where the thermal spectrum
is extracted, which corresponds to only $\sim 0.4$ times the radius
where the X-ray emission is detectable in the {\it Chandra} data.

\subsection{Comparison with dynamical estimates}

Using the virial approach, \citet{cohen2002} derive
$M_{\mathrm{tot}}=(6.92 ^{+2.20} _{-1.89}) \times 10^{14} M_{\sun}$
within 597 kpc from galaxies velocity dispersion measurements.  Within
this radius we consistently find $M_{\mathrm{tot}}=(6.06 \pm 0.61,
6.05 \pm 1.14) \times 10^{14} M_{\sun}$ for model (SO, DDg1).

\subsection{Comparison with gravitational lensing}

Since the gravitational lensing analysis measures the projected total
mass distribution, in order to compare consistently the results from
the X-ray and lensing techniques we project along the line of sight
the cumulative 3D mass profiles $M_{\mathrm{tot}}(<r)$ derived in
Sect. \ref{mass.sec}, thus obtaining
$M_{\mathrm{tot}}^{\mathrm{proj}}(<r)$.

From a weak lensing investigation, \citet{fischer97} derive
$M_{\mathrm{tot}}^{\mathrm{proj}}=(9.35 \pm 2.55) \times 10^{14}
M_{\sun}$ within 850 kpc.  We find, in good agreement,
$M_{\mathrm{tot}}^{\mathrm{proj}}=(9.95 \pm 1.03, 9.17 \pm 2.13.79)
\times 10^{14} M_{\sun}$ for model (SO, DDg1).  Note that
\citet{fischer97} compare their mass measurement
$M_{\mathrm{tot}}^{\mathrm{proj}}$ within 850 kpc with the
$M_{\mathrm{tot}}$ value quoted in \cite{schindler1997} and find a
large discrepancy.  As pointed out by \citet{sahu1998}, the two mass
determinations are in agreement if the correct quantities are
compared.

In the strong lensing analysis by \citet{sahu1998},
$M_{\mathrm{tot}}^{\mathrm{proj}}=5.36 \times 10^{14} M_{\sun}$ is
measured within 204 kpc, the cluster-centric distance of the arcs.
Within this projected distance we find
$M_{\mathrm{tot}}^{\mathrm{proj}}=(2.38 \pm 0.27, 2.05 \pm 0.37)
\times 10^{14} M_{\sun}$ for model (SO, DDg1).  Although this
discrepancy might be due to the fact that we excise the perturbed
region of the cluster, we note that such a large mismatch between the
masses determined from X-rays and strong lensing is commonly found
\citep[see][and references therein]{wu98}.  In the inner core of
clusters, where strong lensing occurs, the physics of the ICM may be
complicated by the interaction with the central AGN.  The central
cluster region is thus poorly described by the usual simple models
used in the X-ray methods which rely on the assumptions of spherical
symmetry and hydrostatic equilibrium.

\begin{figure}
\resizebox{\hsize}{!}{\includegraphics{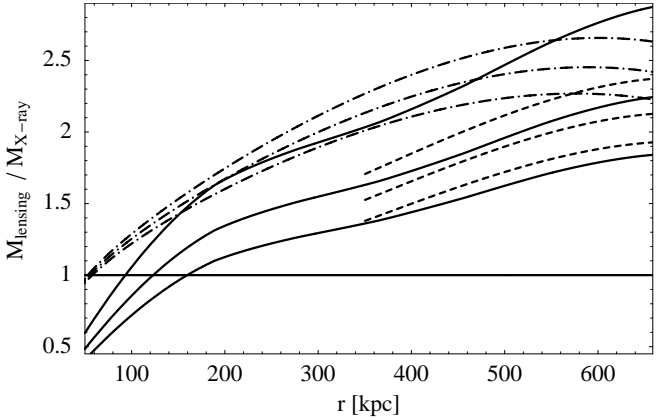}}
\caption{Ratio of the lensing to the (projected) X-ray mass
    profile for different X-ray mass estimates.  The line styles are
    the same as in Fig. \ref{fig:mtot}.  The reported errors are those
    coming from the X-ray mass determinations.}
\label{fig:ratio}
\end{figure}

We compare our total mass determination with the lensing results of
\citet{bradac05}.  The results of \citet{bradac05} are obtained using
a mass reconstruction method which combines strong and weak
gravitational lensing data and effectively breaks the mass-sheet
degeneracy \citep{bradac05method}.  In Fig. \ref{fig:ratio} we show
the X-ray to lensing mass ratio
$M_{\mathrm{lensing}}/M_{\mathrm{X-ray}}$ as a function of radius up
to $\sim 670$ kpc, the limiting radius of the lensing study.  From a
visual inspection of this figure it is clear that there is lack of
agreement between the X-ray and lensing mass estimates.  Only in the
central region the X-ray mass is marginally consistent with the
lensing mass.  The mass ratios increase with radius and tend to
approach a constant value at large radii.  At $600$ kpc the ratio is
2.07, 2.17, and 2.45 for the X-ray mass estimated using the single
$\beta$-model (SO), double $\beta$-model (DDg1), and NFW model,
respectively.  We stress that the same discrepancy is found when we
compare our mass profiles with a corrected mass profile computed from
the lensing map where the SE quadrant, which contains the hot X-ray
subclump, is excluded.

\subsection{Comparison with the Sunyaev-Zel'dovich (SZ) effect}

Through the SZ effect, \citet{etienne2001} measure the gas mass of RX
J1347.  They compare their results with the X-ray results of
\citet{schindler1997}, finding good agreement.  Within 74 arcsec= 427
kpc the SZ estimate is $M_{\mathrm{gas}}= (4.7 \pm 0.4)\times 10^{13}
M_{\sun}$ in agreement with our value, $M_{\mathrm{gas}}=(5.5 \pm
0.1)\times 10^{14} M_{\sun}$.


\section{The cool core}
\label{coolcore.sec}

As shown in Sect. \ref{radial.sec}, there is no evidence for very low
temperature gas in the core of RX J1347, suggesting that the
description of the inner region of this cluster by means of a standard
cooling flow model is not appropriate. The spectral analysis in
\cite{gitti04} shows that if the cool core in RX J1347 is fitted with
an empirical cooling flow model where the lowest temperature is left
as a free parameter, very tight constraints on the existence of a
minimum temperature ($\sim$ 2 keV) are found. This situation is common
for cool core clusters and it has become clear that the gas with short
cooling time at the center of these objects must be prevented from
cooling below the observed central temperature minimum.  The most
appealing mechanism is heating by AGN because it is strongly motivated
by observations.  Central AGNs with strong radio activity are found in
the majority of cool core clusters \citep[e.g,][]{burns90,ball93} and
powerful interactions of the radio sources with the ICM are observed
\citep[e.g.,][and references therein]{birzan,rafferty2006}. The
presence of a central AGN in RX J1347 is indicated by the NRAO VLA Sky
Survey (NVSS), that shows a strong central source along with some hint
of a possible extended emission.  However, the resolution and
sensitivity of the NVSS are not sufficient to study the
characteristics of the central source and establish the existence of
diffuse emission.  We obtained new VLA data in order to further
investigate the nature and properties of the radio source in RX J1347
\citep[][]{gitti07prep}.

In order to explore the heating by AGN, we adopt the model developed
by \citet[][hereafter effervescent heating]{ru02}.  The details of the
model and the procedure adopted to estimate the AGN parameters from
the observed temperature and density profiles are given in
\citet{piff06}.  Here we simply summarize the essential elements.  In
the effervescent heating scenario the central AGN is assumed to inject
buoyant bubbles into the ICM, which heat the ambient medium by doing
$PdV$ work as they rise and expand adiabatically.  In addition,
besides being essential in stabilizing the model, thermal conduction
transports energy from the hotter, outer region to the central region.
Unfortunately its efficiency is poorly known since it depends on
magnetic fields and models with different fractions $f_{\mathrm{c}}$
of the Spitzer rate are studied.  For a fixed $f_{\mathrm{c}}$ between
0 and 1/3 (the maximum for magnetized a plasma) the contribution of
heat conduction as a function of radius is known since the temperature
gradient is estimated from the deprojected temperature profile.  We
note that if one assumes that heat conduction alone balances radiative
losses, then its efficiency would be much larger that 1/3 of the
Spitzer rate and therefore unrealistic. The raising entropy profile
(in Sect. \ref{entropy.sec}) indicates that convection is not
operating on the scales that we are able to resolve and is therefore
not included in the model.  The extra heating profile resulting from
subtracting the heat conduction yield from the ICM emissivity is then
assumed to be balanced by the AGN heating function:
\begin{equation}
 H^{AGN} \propto {L \over {4 \pi r^2}} \left( 1 - e^{-r/r_0} \right) \left({p \over 
p_0}\right)^{(\gamma_b-1)/\gamma_b} \frac 1r {{d\ln{p}}\over {d\ln{r}}} 
\label{H.eq}
\end{equation}
where $p$ is the ICM pressure ($p_0$ some reference value) and
$\gamma_{b}$ the adiabatic index of the buoyant bubbles, which is
fixed to 4/3 (i.e., relativistic bubbles).  Fitting Eq.(\ref{H.eq}) to
the extra heating profile provides the AGN parameters $L$ (the {\it
  time-averaged} luminosity) and $r_0$ (the scale radius where the
bubbles start rising in the ICM).  Only if $0.10 \leq f_c \leq 0.27$
the fitting provides meaningful results.  For $f_c=0.27$ the AGN
parameters are $L=7.45 \times 10^{45} \, \mathrm{erg} \,
\mathrm{s}^{-1}$ and $r_\mathrm{0}=4$ kpc.  As we decrease $f_c$ both
AGN parameters increase monotonically and reach the maximum at
$f_c=0.10$, for which we find $L=10.11 \times 10^{45} \, \mathrm{erg}
\, \mathrm{s}^{-1}$ and $r_\mathrm{0}=29$ kpc.  The trend of the AGN
parameters with $f_c$ indicates that, in the framework of the
effervescent heating scenario, heat conduction and AGN heating
cooperate in quenching radiative cooling.  The inferred AGN {\it
  time-averaged} luminosity lies therefore in a quite small range
($7.45-10.11 \times 10^{45} \, \mathrm{erg} \,\mathrm{s}^{-1}$), and
is larger but comparable to the cluster luminosity in the energy range
[2.0-10.0] keV ($L_X = 6.2 \pm 0.2 \times 10^{45} \mbox{ erg
  s}^{-1}$).  The model with $f_c=0.22$ is the one with the smallest
reduced $\chi^2$ and in this case $L=8.32 \times 10^{45} \,
\mathrm{erg} \,\mathrm{s}^{-1}$ and $r_\mathrm{0}=13$ kpc.

The effervescent heating model applied to RX J1347 predicts that the
scale where the bubbles start rising in the ICM is in the range 4-29
kpc. The observed extension of the AGN jets should be of the same
order of magnitude \citep{piff06}. Interestingly, the first results
from 1.4 GHz VLA observations of the central region of RX J1347 show
hints of faint structures emanating from the discrete radio source out
to $\sim 20$ kpc from the center \citep[][]{gitti07prep}. A comparison
between the derived luminosity $L$ with the observed AGN luminosity is
unfortunately not possible. In fact, in the framework of the
effervescent heating model, the derived AGN luminosity is a
time-averaged {\it total} AGN power and a fair comparison is possible
only if the {\it total} jet power is estimated (X-ray and radio powers
are known to be poor tracers of the total AGN power). At present, this
was done only for M87 in the Virgo cluster \citep{owen00}.


\section{Summary}

As indicated by previous studies \citep{allen02b,etienne2001,gitti04},
the cluster RX J1347 shows both the signatures of strong cooling flow
and subcluster merger, that are rarely observed in the same system.

We analyze the data excluding the SE quadrant, where the presence of a
hot X-ray subclump is suggesting that a minor merger has recently
occurred or is still going on, and find that:
\begin{itemize}
\item the features (shape, normalization, scaling properties) of
  density, temperature, entropy, and cooling time profiles are fully
  consistent with those of relaxed, cool core clusters, with no
  indications of perturbations that may originate from the disturbed
  region of the cluster.
\end{itemize}

The usual assumptions of hydrostatic equilibrium and spherical
symmetry can therefore be adopted when analyzing the data with the SE
quadrant excluded. This allows us to perform a detailed mass
reconstruction by starting from the temperature and density profiles
derived from the X-ray analysis. We find that:
\begin{itemize}
\item the total mass profiles computed from a double and single
  $\beta$-model for the surface brightness give consistent results if
  the cool core is excised in the latter case;
\item there is a reasonably good agreement between the total mass
  profile estimated from a double $\beta$-model and from the
  assumption of a NFW profile.  The differences between these
  estimates might come from a poor spatial resolution of the density
  and temperature profiles in the central region, which could bias the
  NFW method;
\item the characteristic radii $r_\mathrm{{\Delta}}$ computed from the
  mass profile are in agreement with the observed size-temperature
  relation, although they are systematically lower than those derived
  by calibrating the relation with numerical simulations.  The mass
  estimated from the NFW profile is in agreement with the observed
  mass-temperature relation, whereas that derived from the double
  $\beta$-model profile is a factor $\sim 1.7$ higher.
\end{itemize}

We compare our gas and total mass estimates with previous work and find that:
\begin{itemize}
\item our estimates of gas and total mass are generally in good
  agreement with those from previous X-ray, dynamical, weak lensing
  and SZ studies;
\item a discrepancy of a factor $\sim 2$ between strong lensing and
  X-ray mass determinations is confirmed;
\item there is a large discrepancy at all radii between our total mass
  estimate and the mass reconstructed through the combination of both
  strong and weak lensing.
\end{itemize} 
 
We study the AGN heating in RX J1347 by applying the effervescent
heating model. We find support to the picture that AGN heating and
heat conduction cooperate in balancing radiative losses. Our
predictions concerning the extension of the AGN jets in RX J1347 are
consistent with recent radio observations of the radio source at the
cluster center.


\begin{acknowledgements}
  This work is based on observations obtained with
  \textit{XMM--Newton}, an ESA science mission with instruments and
  contributions directly funded by ESA Member States and the USA
  (NASA).  We thank the referee for insightful comments which improved
  the presentation of the results.  We thank M. Brada{\v c} for
  providing the lensing results.  MG thanks S. Ettori and F. Brighenti
  for useful discussions.  MG acknowledges the support by Austrian
  Science Foundation FWF grant P15868, by Grant NNG056K87G from NASA's
  Goddard Space Flight Center and by NASA Long Term Space Astrophysics
  Grant NAG4-11025.  RP acknowledges support from the Tiroler
  Wissenschaftsfond (Gef\"ordert aus Mitteln des vom Land Tirol
  eingerichteten Wissenschaftsfonds).  SS aknowledges support from the
  Austrian Science Foundation FWF under grant P18SL3-N1G.

\end{acknowledgements}



\end{document}